\documentclass{article}
\usepackage{spconf,amsmath,graphicx}
\usepackage{enumitem}
\setlist{nosep, leftmargin=14pt}

\usepackage{mwe} 

\usepackage{algorithm}
\usepackage{algorithmicx}
\usepackage{algpseudocode}
\usepackage{amsmath}
\usepackage{times}
\usepackage{bbm}
\usepackage{helvet}
\usepackage{courier}
\usepackage{mathtools,xparse}
\usepackage{array}
\usepackage{graphicx}
\usepackage{multicol}
\usepackage{multirow}
\usepackage[flushleft]{threeparttable}

\title{T-Net: Learning Feature Representation with Task-specific Supervision for Biomedical Image Analysis}
\name{Weinan Song, Yuan Liang, Jiawei Yang, Kun Wang, and Lei He}
\address{Design Automation Laboratory, University of California, Los Angeles, USA}

\begin{document}

\maketitle
\begin{abstract}
The encoder-decoder network is widely used to learn deep feature representations from pixel-wise annotations in biomedical image analysis. Under this structure, the performance profoundly relies on the effectiveness of feature extraction achieved by the encoding network. However, few models have considered adapting the attention of the feature extractor even in different kinds of tasks. In this paper, we propose a novel training strategy by adapting the attention of the feature extractor according to different tasks for effective representation learning. Specifically, the framework, named T-Net, consists of an encoding network supervised by task-specific attention maps and a posterior network that takes in the learned features to predict the corresponding results. The attention map is obtained by the transformation from pixel-wise annotations according to the specific task, which is used as the supervision to regularize the feature extractor to focus on different locations of the recognition object. To show the effectiveness of our method, we evaluate T-Net on two different tasks, \textit{i.e.}, segmentation and localization. Extensive results on three public datasets (BraTS-17, MoNuSeg and IDRiD) have indicated the effectiveness and efficiency of our proposed supervision method, especially over the conventional encoding-decoding network.

\end{abstract}
%
%
\section{Introduction}
Feature extraction plays a vital role in image analysis. With the rapid advance of deep learning, the feature extractor can be learned automatically by training a deep neural network (DCNN) on large-scale labelled images. However, as most biomedical image datasets are notoriously small due to the tedious labelling work and professional knowledge requirement, many models take an encoding-decoding network as the backbone to learn feature representations from pixel-wise annotations. This encoder-decoder architecture generally consists of a symmetrical encoding and decoding network to learn a non-linear mapping function between the image and the annotations. Moreover, the feature extractor, played by the encoding network, can be further exploited in other tasks, such as segmentation or detection. For example, Y-Net \cite{YNet} takes the encoding-decoding network to learn joint information of segmentation and classification for diagnosis of breast biopsy images. Retina U-Net \cite{RetinaUNet} utilizes the encoder-decoder network to explore segmentation supervision for lesion detection in lung CT and breast MRI images.

However, under such an encoder-decoder structure, most models follow the same way to train the feature extractor, where the encoding network is trained together with the decoding network by pixel-wise annotations. Therefore, the feature extractor is hard to adjust its attention during the training process, even when the model only cares about part of the recognition object. For example, in a detection task, the model should pay more attention to the object center instead of equal attention to different locations of the detection target. To solve this, we propose a new supervision method to optimize feature representations of the encoding network according to the specific task, which can adjust attention to accommodate the feature extractor to different situations.

\begin{figure}
    \centering
    \includegraphics[width=0.45\textwidth]{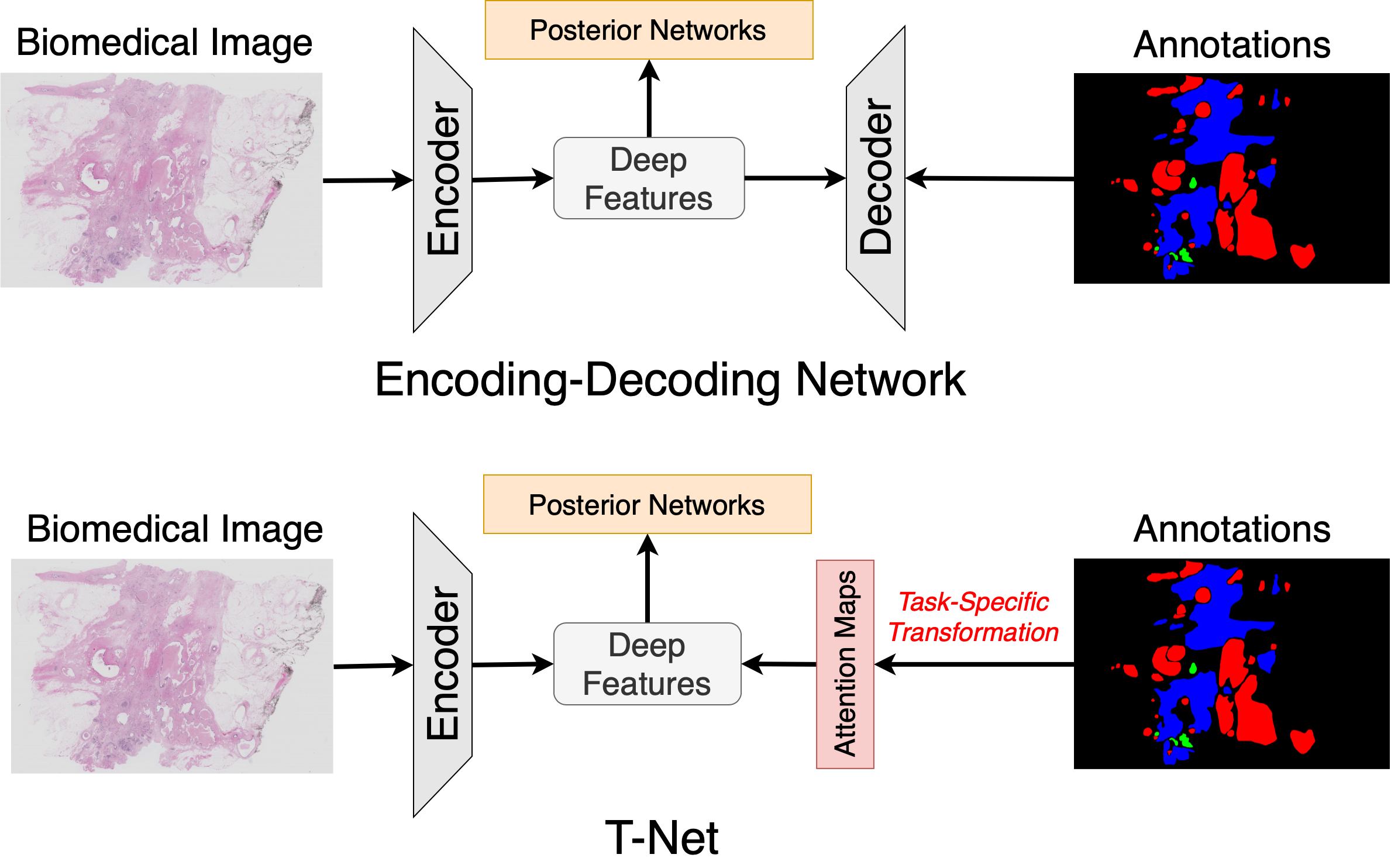}
    \caption{We compare T-Net with a general encoding-decoding network under a biomedical analysis framework in this figure. The analysis task module takes features generated by the encoding network as input for different predictions, such as coordinates in a detection task. In the encoding-decoding network, the feature extractor is trained directly from pixel-wise annotations. However, in T-Net, we directly supervise the encoding network with attention maps generated by task-specific transformations from the annotation.}
    \label{fig:overview}
\end{figure}

As shown in Fig. \ref{fig:overview}, our proposed framework, named T-Net, consists of an encoding network to directly learn feature representations from task-specific attention maps and a posterior network to predict the results from the learned features. Specifically, the attention maps are obtained by various transformations from the binary mask of each recognition object according to the task. Furthermore, the posterior network is trained by well-learned features generated by the encoding network to predict the corresponding output, such as binary masks for segmentation or coordinates for object detection. To show the effective feature extraction of our method, we evaluate T-Net on three public datasets for comparison with state-of-the-art methods, where T-Net achieves comparable or superior performance to the leading methods on the open leaderboard. Furthermore, we also set an ablation study to show the impact of different attention maps, which indicates a significant advantage in feature learning over the conventional encoding-decoding network.
\section{Methodologies}

\subsection{Supervision from the Attention Map}
\subsubsection{Optimization of the Feature Extractor}
In a general encoding-decoding network, feature extraction is learned by the encoding network from binary masks $I^{B}$ of recognition objects. If we take a binary-class recognition problem, for example, the optimization problem of training an encoding network ${f_e}$ can be concluded as: 

\begin{equation}
    \label{eq:Opt_UNet}
    \mathop {\min }\limits_{{\theta _e},{\theta _d}} Loss({f_d}({\theta _{d,}}{f_e}({\theta _e},I)),I^{B}),
\end{equation}
where $I$ and ${f_d}$ represent the input image and the decoding network. 
Under this structure, the optimization process of the feature extractor relies on the back-propagated gradient from the decoding network and the supervision format is fixed even for different tasks. However, in T-Net, we directly train the encoding network by deep supervision with task-specific attention maps. Then the optimization problem for the feature extractor can be changed to:
\begin{equation}
\label{eq:Opt_TNet}
\begin{split}
    \mathop {\min }\limits_{{\theta _e}} Loss({f_e}({\theta _e},I),Trans(I^{B})),
\end{split}
\end{equation}
where $Trans()$ is the task-specific transformation function.
Compared with the encoding-decoding network, T-Net can adjust its attention in feature representations when training the encoding network by utilizing different attention maps, leading to an improvement in the overall performance in different tasks for biomedical image analysis.

\subsubsection{Loss Function of the Feature Extractor}
As the recognition object usually takes only a small part of the whole image, the loss function should consider the large background area. Therefore, we use dice loss to optimize the encoding network with $N$ ${H \times W}$ attention maps as:
\begin{equation}
\label{eq:Template_loss}
Loss = 1 - \frac{1}{{{N}}}\sum\limits_t^{{N}} {\frac{{2\sum\limits_i^{{H}} {\sum\limits_j^{{W}} {{p_{i,j,t}}{g_{i,j,t}}} } }}{{\sum\limits_i^{{H}} {\sum\limits_j^{{W}} {p_{i,j,t}^2}  + \sum\limits_i^{{H}} {\sum\limits_j^{{W}} {g_{i,j,t}^2} } } }}},
\end{equation}
where $(p_{i,j,t}, g_{i,j,t})$ represent the prediction and ground truth point pair for the $t$-th attention map.

\subsection{Generation of the Attention Map}
The attention map \cite{AtentionMap} has been widely used in semantic segmentation models to improve the learning ability by establishing connections between different locations in the feature map. In T-Net, we utilize the attention map in a different way by taking it as deep supervisions to train the feature extractor. In this paper, we propose three ways to obtain the attention maps from pixel-wise annotations, enabling the encoding network to pay attention to different areas of the recognition target. Examples of different attention-maps used in this paper can be seen in Fig. \ref{fig:TNet_Figure}. To be noted, the template generation is not limited to these three methods. We can also deploy other types of attention maps for supervision, depending on applications.

\subsubsection{Shape-Aware Attention Map}
The basic attention map should play the same role as in a conventional encoder-decoder network, where the feature extractor focuses merely on the object shape. Therefore, we directly down-sample the binary mask of the recognition target with nearest-neighbour interpolation, which can guarantee each pixel value in the attention map is either 0 or 1.

\subsubsection{Contour-Aware Attention Map}
To make the encoding network pay additional attention to the area outside the boundary of targets, we expand the shape-aware attention map by applying a normalized Gaussian filter. The Gaussian filter can make a smooth transition from foreground to background on the object boundary and the normalization can guarantee the pixel value is between 0 and 1.

\subsubsection{Center-Aware Attention Map}
To encourage the encoding network to focus on the object center, we transform the shape-aware attention map into a normalized distance map, where each pixel value inside the object reflects the distance between it and the object boundary. For normalization, we divide the distance map by the largest distance value, which should belong to the center point as it is most far away from the background.

\begin{figure}[tp]
    \centering
    \includegraphics[width=0.45\textwidth]{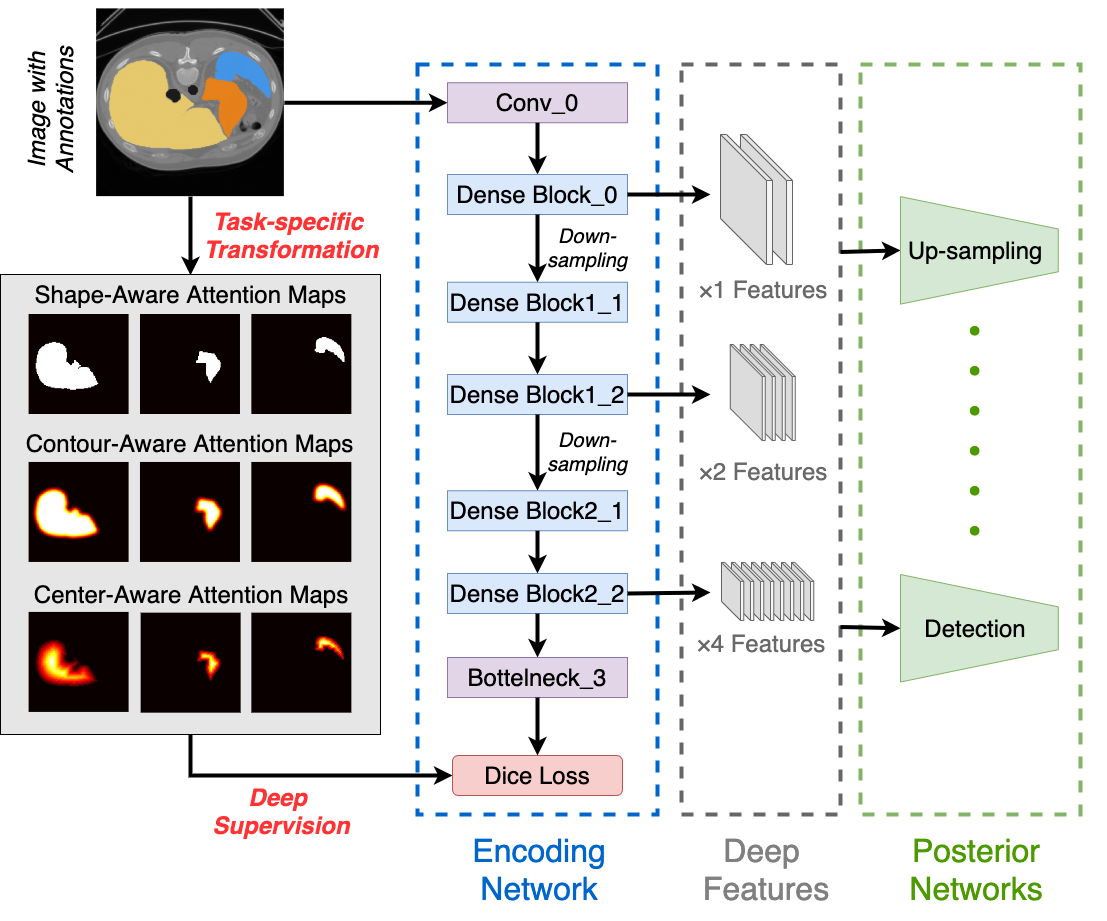}
    \caption{We show the detailed structure of T-Net in this picture, where the framework consists of an encoding network to learn feature representations and posterior networks for different analysis tasks. 
    }
    \label{fig:TNet_Figure}
\end{figure}

\subsection{Network Architecture}
In this section, we introduce the detailed architecture of T-Net. As shown in Fig. \ref{fig:TNet_Figure}, the framework consists of an encoding network to learn feature representations and multiple posterior networks for different analysis tasks. The encoding network is optimized by attention maps generated by different methods according to the specific task, and the posterior networks are trained with multi-level features generated from the encoding network.

\subsubsection{Encoding Network}
We use dense block \cite{DenseConv} as our backbone to learn different levels of deep features in the encoding network. As shown in the blue box in Fig. \ref{fig:TNet_Figure}, the encoding network has two steps of down-sampling to extract features. The $convolution$ layer on the top maps the input image into fixed-number feature maps. Then $\times1$ features are generated from the first dense block, while $\times2$ and $\times4$ features are obtained from two consecutive dense blocks after a $max-pooling$ layer. Finally, the $\times4$ features are reduced to $N$ channels by a $bottleneck$ layer, which consists of an $1\times1 convolution$ layer and a $Sigmoid$ layer.

\subsubsection{Posterior Networks}
The architecture of a posterior network depends on the kind of task in image analysis. For example, the posterior network should be an up-sampling network for segmentation and regression network for detection. To be noted, under a segmentation task, T-Net is still different from a general encoding-decoding network as the two networks in T-Net are trained separately.


\section{Experiments}

\begin{figure*}[tp]
    \centering
    \label{fig:compare}
    \includegraphics[width=\textwidth]{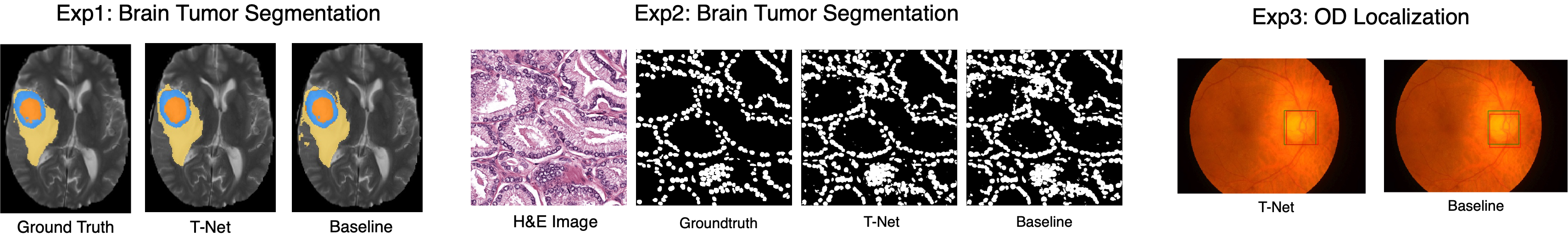}
    \caption{In this figure, we show example results for T-Net and the baseline network in three experiments. With different attention maps as the supervision, the encoding network can accommodate the feature representations to the specific task, leading to an improvement in the overall performance.}
    \label{fig:Resutls_Compare}
\end{figure*}

To prove the effectiveness of our proposed supervision method, We evaluate T-Net on three public datasets for three different tasks: (1) the BraTS-17 dataset \cite{BraTS1} \cite{BRATS2} for 3D brain tumor segmentation, (2) the MoNuSeg dataset \cite{MoNuSeg} for 2D semantic nuclei segmentation, and (3) the IDRid dataset \cite{IDRiD1} for optic disc (OD) localization.

\subsection{Experiment Settings}
\subsubsection{Evaluation Metrics}
For tumor segmentation, we use dice score (Dice) and 95\% Hausdorff distance (Hausdorff) to evaluate the segmentation performance over different areas of brain tumor, \textit{i.e.}, the whole tumor (WT),the tumor core (TC) and the enhancing tumor (ET). For convenience, we also use a averaged score $S = \sum\nolimits_{class} {(Dice/200 - Hausdorff/60)}$, similar to \cite{Ensemble}, to compare the overall performance.For nuclei segmentation, we follow the same way as in \cite{MoNuSeg} to split the dataset into two categories, where the model is train only in the first group. The semantic segmentation performance is evaluated by average dice coefficient in each group. For OD localization, we use the images with pixel-level annotations to train the feature extractor and the images with instance-level marks of OD to train the detection network. The localization performance is evaluated over the averaged Euclidean distance (ED).

\subsubsection{Selection of Attention Maps}
For tumor segmentation, to distinguish the tentacle-like tumor and its surroundings, we use shape-aware attention maps for each tumor area to train the encoding network. For nuclei segmentation, we use the contour-aware attention maps to extend attention of the nuclei boundary. For OD localization, we utilize center-aware attention maps with Chebyshev distance to encourage the feature extractor focus on the OD center.

\subsubsection{Network Settings}
For dense block in the encoding network, we set $K=8$ and $N=4$. The channel number of multi-level features are 16, 32, and 64. For tumor and nuclei segmentation, we use a symmetrical 3D/2D up-sampling network as the posterior network. For OD localization, we use a similar detection network as in \cite{YOLO9000} to predict coordinates. All the networks are optimized by Adam with a learning rate of $10^{-3}$.

\begin{table}[tp]
    \small
    \renewcommand{\arraystretch}{1.2}
    \caption{Evaluation on Brain Tumor Segmentation.}
    \label{tab:BRATS_compare}
    \centering
    \setlength\tabcolsep{3pt}
    \begin{tabular}{p{2.8cm}<{\centering}p{0.5cm}<{\centering}p{0.5cm}<{\centering}p{0.5cm}<{\centering}p{0.1cm}p{0.5cm}<{\centering}p{0.5cm}<{\centering}p{0.5cm}<{\centering}p{0.1cm}p{0.5cm}<{\centering}}
    \hline
    \multirow{2}{*}{Method}&\multicolumn{3}{c}{Dice}&&\multicolumn{3}{c}{Hausdorff}&&\multirow{2}{*}{S}\cr\cline{2-4}\cline{6-8}
    &WT&TC&ET&&WT&TC&ET&\cr
    \hline
    Sequential U-Net \cite{Seq_U-Net}&88.2&73.2&73.0&&8.12&11.4&6.17&&0.74\cr
    3D FCN \cite{3DFCN}&89.9&75.1&71.3&&4.16&8.65&6.98&&0.85\cr
    Residual U-Net \cite{ResUNet}&89.6&79.7&73.2&&6.97&9.48&4.55&&0.86\cr
    T-Net&88.9&76.7&71.5&&4.86&\textbf{8.20}&\textbf{4.46}&&\textbf{0.89}\cr
    \hline
    \end{tabular}
\end{table}

\begin{table}[tp]
    \small
    \renewcommand{\arraystretch}{1.2}
    \caption{Evaluation on Multi-organ Nuclei Segmentation.}
    \label{tab:MoNuSeg_compare}
    \centering
    \begin{tabular}{p{2.8cm}<{\centering}p{1.2cm}<{\centering}p{1.2cm}<{\centering}p{1.2cm}<{\centering}}
    \hline
    Method&Test A&Test B&Overall\cr
    \hline
    Dist U-Net \cite{DistMap}&0.7756&0.8005&0.7863\cr
    RIC U-Net \cite{RICUNet}&-&-&0.8008\cr
    FullNet \cite{FullNet}&0.8007&0.8054&0.8027\cr
    T-Net &0.7930&\textbf{0.8346}&\textbf{0.8108}\cr
    \hline
    \end{tabular}
\end{table}

\subsection{Results}
\subsubsection{Comparison with State-of-the-art Methods}
We first compare T-Net with other methods and show the results in Table \ref{tab:BRATS_compare}-\ref{tab:IDRid_compare}, where T-Net achieves the best performance by task-specific supervisions. For tumour segmentation, T-Net achieves little improvement for overall performance over Residual U-Net, the leading single-model method, mainly because the shape-aware attention maps play the same role as general deep supervisions. For nuclei segmentation, the improvement, especially in TestB, mainly comes from the additional attention to the nuclei boundary. For OD localization, the posterior network can detect the OD center more accurately when the feature extractor focuses on the object center.

\subsubsection{Comparison with different Supervisions}
To further validate our proposed supervision method, we utilize the three proposed attention maps in all three experiments to show the influence of selecting different attention maps. Furthermore, we also use a baseline encoding-decoding network with the same architecture as T-Net but without deep supervision to alleviate the impact of the network architecture. As shown in Table \ref{tab:TNet_compare} and Fig. \ref{fig:compare}, T-Net with proper supervisions can significantly outperform the baseline network. While the performance may also decrease if the model is supervised by wrong attention maps, especially when the model pays too much attention to unimportant locations.

\begin{table}[tp]
    \small
    \renewcommand{\arraystretch}{1.2}
    \caption{Evaluation for OD Localization.}
    \label{tab:IDRid_compare}
    \centering
    \begin{tabular}{p{4cm}<{\centering}p{2cm}<{\centering}}
    \hline
    Method&ED\cr
    \hline
    RPI-based faster RCNN \cite{OD_RCNN}&32.60\cr
    Relation Net \cite{RelationNet}&26.12\cr
    DeepDR&25.62\cr
    T-Net &\textbf{24.85}\cr
    \hline
    \end{tabular}
\end{table}

\begin{table}[tp]
    \small
    \renewcommand{\arraystretch}{1.2}
    \caption{Comparison of T-Net and the baseline.}
    \label{tab:TNet_compare}
    \centering
    \begin{tabular}{p{2cm}<{\centering}p{1.2cm}<{\centering}p{1cm}<{\centering}p{1.2cm}<{\centering}p{1cm}<{\centering}}
    \hline
    Attention Map&BraTS&MoNuSeg&IDRid\cr
    \hline
    -&0.85&0.7987&32.61\cr
    Shape-Aware&\textbf{0.89}&0.8028&42.10\cr
    Contour-Aware&0.76&\textbf{0.8108}&34.21\cr
    Center-Aware&0.83&0.8066&\textbf{24.85}\cr
    \hline
    \end{tabular}
\end{table}

\section{Conclusion}
In this paper, we propose a new supervision method to effectively learn deep representations for an encoding network by adapting attention of the feature extractor according to the specific task. We utilize three types of attention maps to deeply supervised the encoding network according to the specific task. Extensive experiments on three public datasets have shown that our training strategy is simple yet effective and can be applied to various tasks.



\clearpage
\bibliographystyle{IEEEbib}
\bibliography{refs}

\begin{thebibliography}{10}

\bibitem{YNet}
Sachin Mehta, Ezgi Mercan, Jamen Bartlett, Donald Weaver, Joann~G Elmore, and
  Linda Shapiro,
\newblock ``Y-net: joint segmentation and classification for diagnosis of
  breast biopsy images,''
\newblock in {\em International Conference on Medical Image Computing and
  Computer-Assisted Intervention}. Springer, 2018, pp. 893--901.

\bibitem{RetinaUNet}
Paul~F Jaeger, Simon~AA Kohl, Sebastian Bickelhaupt, Fabian Isensee,
  Tristan~Anselm Kuder, Heinz-Peter Schlemmer, and Klaus~H Maier-Hein,
\newblock ``Retina u-net: Embarrassingly simple exploitation of segmentation
  supervision for medical object detection,''
\newblock in {\em Machine Learning for Health Workshop}. PMLR, 2020, pp.
  171--183.

\bibitem{AtentionMap}
Kunpeng Li, Ziyan Wu, Kuan-Chuan Peng, Jan Ernst, and Yun Fu,
\newblock ``Tell me where to look: Guided attention inference network,''
\newblock in {\em Proceedings of the IEEE Conference on Computer Vision and
  Pattern Recognition}, 2018, pp. 9215--9223.

\bibitem{DenseConv}
Gao Huang, Zhuang Liu, Laurens Van Der~Maaten, and Kilian~Q Weinberger,
\newblock ``Densely connected convolutional networks,''
\newblock in {\em Proceedings of the IEEE conference on computer vision and
  pattern recognition}, 2017, pp. 4700--4708.

\bibitem{BraTS1}
Bjoern~H Menze, Andras Jakab, Stefan Bauer, Jayashree Kalpathy-Cramer, Keyvan
  Farahani, Justin Kirby, Yuliya Burren, Nicole Porz, Johannes Slotboom, Roland
  Wiest, et~al.,
\newblock ``The multimodal brain tumor image segmentation benchmark (brats),''
\newblock {\em IEEE transactions on medical imaging}, vol. 34, no. 10, pp.
  1993--2024, 2015.

\bibitem{BRATS2}
Spyridon Bakas, Hamed Akbari, Aristeidis Sotiras, Michel Bilello, Martin
  Rozycki, Justin~S Kirby, John~B Freymann, Keyvan Farahani, and Christos
  Davatzikos,
\newblock ``Advancing the cancer genome atlas glioma mri collections with
  expert segmentation labels and radiomic features,''
\newblock {\em Scientific data}, vol. 4, pp. 170117, 2017.

\bibitem{MoNuSeg}
Neeraj Kumar, Ruchika Verma, Sanuj Sharma, Surabhi Bhargava, Abhishek Vahadane,
  and Amit Sethi,
\newblock ``A dataset and a technique for generalized nuclear segmentation for
  computational pathology,''
\newblock {\em IEEE transactions on medical imaging}, vol. 36, no. 7, pp.
  1550--1560, 2017.

\bibitem{IDRiD1}
Prasanna Porwal, Samiksha Pachade, Ravi Kamble, Manesh Kokare, Girish Deshmukh,
  Vivek Sahasrabuddhe, and Fabrice Meriaudeau,
\newblock ``Indian diabetic retinopathy image dataset (idrid): a database for
  diabetic retinopathy screening research,''
\newblock {\em Data}, vol. 3, no. 3, pp. 25, 2018.

\bibitem{Ensemble}
Hao Zheng, Yizhe Zhang, Lin Yang, Peixian Liang, Zhuo Zhao, Chaoli Wang, and
  Danny~Z Chen,
\newblock ``A new ensemble learning framework for 3d biomedical image
  segmentation,''
\newblock in {\em Proceedings of the AAAI Conference on Artificial
  Intelligence}, 2019, vol.~33, pp. 5909--5916.

\bibitem{YOLO9000}
Joseph Redmon and Ali Farhadi,
\newblock ``Yolo9000: better, faster, stronger,''
\newblock in {\em Proceedings of the IEEE conference on computer vision and
  pattern recognition}, 2017, pp. 7263--7271.

\bibitem{Seq_U-Net}
Andrew Beers, Ken Chang, James Brown, Emmett Sartor, CP~Mammen, Elizabeth
  Gerstner, Bruce Rosen, and Jayashree Kalpathy-Cramer,
\newblock ``Sequential 3d u-nets for biologically-informed brain tumor
  segmentation,''
\newblock {\em arXiv preprint arXiv:1709.02967}, 2017.

\bibitem{3DFCN}
Andrew Jesson and Tal Arbel,
\newblock ``Brain tumor segmentation using a 3d fcn with multi-scale loss,''
\newblock in {\em International MICCAI Brainlesion Workshop}. Springer, 2017,
  pp. 392--402.

\bibitem{ResUNet}
Fabian Isensee, Philipp Kickingereder, Wolfgang Wick, Martin Bendszus, and
  Klaus~H Maier-Hein,
\newblock ``Brain tumor segmentation and radiomics survival prediction:
  contribution to the brats 2017 challenge,''
\newblock in {\em MICCAI Brainlesion Workshop}. Springer, 2017, pp. 287--297.

\bibitem{DistMap}
Peter Naylor, Marick La{\'e}, Fabien Reyal, and Thomas Walter,
\newblock ``Segmentation of nuclei in histopathology images by deep regression
  of the distance map,''
\newblock {\em IEEE transactions on medical imaging}, vol. 38, no. 2, pp.
  448--459, 2019.

\bibitem{RICUNet}
Zitao Zeng, Weihao Xie, Yunzhe Zhang, and Yao Lu,
\newblock ``Ric-unet: An improved neural network based on unet for nuclei
  segmentation in histology images,''
\newblock {\em IEEE Access}, vol. 7, pp. 21420--21428, 2019.

\bibitem{FullNet}
Hui Qu, Zhennan Yan, Gregory~M Riedlinger, Subhajyoti De, and Dimitris~N
  Metaxas,
\newblock ``Improving nuclei/gland instance segmentation in histopathology
  images by full resolution neural network and spatial constrained loss,''
\newblock in {\em International Conference on Medical Image Computing and
  Computer-Assisted Intervention}. Springer, 2019, pp. 378--386.

\bibitem{OD_RCNN}
Xuechen Li, Linlin Shen, and Jiang Duan,
\newblock ``Optic disc and fovea detection using multi-stage region-based
  convolutional neural network,''
\newblock in {\em Proceedings of the 2nd International Symposium on Image
  Computing and Digital Medicine}, 2018, pp. 7--11.

\bibitem{RelationNet}
Sudharshan~Chandra Babu, Shishira~R Maiya, and Sivasankar Elango,
\newblock ``Relation networks for optic disc and fovea localization in retinal
  images,''
\newblock {\em arXiv preprint arXiv:1812.00883}, 2018.

\end{thebibliography}

\end{document}